\documentclass[9pt,aps,preprintnumbers,showkeys]{revtex4}
\usepackage[utf8]{inputenc}
\usepackage{graphicx}
\usepackage{amssymb}
\usepackage{bm}
\usepackage[sumlimits,intlimits]{amsmath}
\usepackage{amsfonts,amssymb}
\usepackage{amsmath}
\usepackage{bm}
\usepackage{graphics}
\usepackage{graphicx}
\usepackage{textcomp}
\usepackage{color}
\usepackage{multirow}

\def\L{{\cal L}}

\def\L{{\cal L}}
\def\sech{\,\mbox{sech}}

\begin{document}
\title{Supersymmetric Quantum Mechanics: two factorization schemes, and quasi-exactly solvable potentials.}
\author{J. Socorro}
\email{socorro@fisica.ugto.mx}
\affiliation{Departamento de  F\'{\i}sica, DCeI, Universidad de Guanajuato-Campus Le\'on, C.P. 37150, Le\'on, Guanajuato, M\'exico}
\author{Marco A. Reyes}
\email{marco@fisica.ugto.mx}
\affiliation{Departamento de  F\'{\i}sica, DCeI, Universidad de Guanajuato-Campus Le\'on, C.P. 37150, Le\'on, Guanajuato, M\'exico}
\author{Carlos Villaseñor Mora}
\email{vimcarlos@fisica.ugto.mx}
\affiliation{Departamento de  F\'{\i}sica, DCeI, Universidad de Guanajuato-Campus Le\'on, C.P. 37150, Le\'on, Guanajuato, M\'exico}
\author{Edgar Condori Pozo}
\email{edgarcondoripozo@gmail.com}
\affiliation{Departamento de  F\'{\i}sica, DCeI, Universidad de Guanajuato-Campus Le\'on, C.P. 37150, Le\'on, Guanajuato, M\'exico}

\begin{abstract}
We present the general ideas on SuperSymmetric Quantum Mechanics (SUSY-QM) using different representations for the operators in question, which are defined by the corresponding bosonic Hamiltonian as part of SUSY Hamiltonian and its supercharges, which are defined as matrix or differential operators.
We show that, although most of the SUSY partners of one-dimensional Schr\"odinger problems have already been
found,there are still some unveiled aspects of the factorization procedure which may lead to richer insights of the problem involved.
\end{abstract}


\keywords{Supersymmetric quantum mechanics, Quasi-exactly solvable potentials}

\maketitle

\section{Introduction}

We present the general ideas on SuperSymmetric Quantum Mechanics (SUSY-QM) using different representations for the operators in question, which are defined by the corresponding bosonic Hamiltonian as part of SUSY Hamiltonian and its supercharges, $\rm \hat Q^-$ and $\rm \bar Q^+$, which are defined as matrix or differential operators.
We show that, although most of the SUSY partners of one-dimensional Schr\"odinger problems have already been
found,\cite{cooper} there are still some unveiled aspects of the factorization procedure which may lead to richer insights of the problem involved.
In particular, we refer to the factorization of the Hamiltonian in terms of two non-mutually-adjoint operators.\cite{ranferi,rafael}

In this work we try three main schemes, the first one consists on finding the
eigenvalue  Schrodinger equation in one dimension
using the matrix  representation via the appropriate factorization
with  ladder like operators, and finding the one parameter isospectral equation for
this one. In this scheme the wave
function is written as a supermultiplet. Continuining with the
Schrodinger model, we extend SUSY to include two parameters
factorizations, which include the SUSY factorization as particular
case. As examples, we include the case of the harmonic oscillator and
the P\"oschl-Teller potentials. Also, we include the steps for the
two-dimensional case and apply it to particular cases. The
second scheme uses the differential representation in
Grassmann numbers, where the wave function can be written as an
n-dimensional vector or as an expansion in Grassmann variables
multiplied by bosonic functions. We apply the scheme in two bosonic
variables a particular cosmological model and compare the corresponding
solutions found.
The third scheme trias on extensions to the SUSY factorization, and to the case of
quasi-exactly solvable potentials;  we present a particular case
which does not form part of the class of potentials found using Lie
algebras.

To establish the different approaches presented here, we will briefly describe the different main formalisms applied to supersymmetric quantum mechanics, techniques that are now widely used
 in a rich spectrum of physical problems, cover such diverse fields as particle physics, quantum field theory, quantum gravity, quantum cosmology and statistical mechanics, to mention some of them:

\begin{itemize}
\item{} In one dimension, SUSY-QM may be considered an equivalent formulation of the Darboux transformation method, which is well known in mathematics from the original paper of Darboux \cite{darboux}, the book  by Ince \cite{Ince}, and the book by Matweev and Salle \cite{MS}, where the method is widely used in the context of the soliton theory. An essential ingredient of the method, is the particular choice of a transformation operator in the form of a differential operator which intertwines two Hamiltonian and relates their eigenfunctions. When this approach is applied to quantum theory, it allows  to generate a huge family of exactly solvable local potential starting with a given exactly solvable local potential \cite{CKS}. This technique is also known in the literature as isospectral formalism, \cite{Mielnik,Nieto,Fernandez,CKS}.
\item{} Those defined by means of the use of supersymmetry as a square root \cite{BG,OSB,lidsey,sm}, in which the Grassmann variables are auxiliary variables and are not  identified as the supersymmetric partners of the bosonic variables. In this formalism, a differential representation is used for the Grassmann variables. Also the  supercharges for the n-dimensional case  read as
\begin{equation}
\rm \hat Q^- = \psi^\mu \left[-\hbar \partial_{q^\mu} + \frac{\partial
S}{\partial q^\mu} \right], \qquad \rm \hat Q^+ = \bar \psi^\nu
\left[-\hbar \partial_{q^\nu} - \frac{\partial S}{\partial q^\nu}
\right], \label{supercharge2}
\end{equation}
where  $\rm S$ is known as the super-potential function which are related to the physical potential under consideration, when the hamiltonian density is written as the Hamilton-Jacobi equation,
 and the  algebra for the variables  $\psi^\mu$ and $\bar \psi^\nu$ is,
\begin{equation}\rm
\left\{ \psi^\mu ,\bar \psi^\nu \right \} = \eta^{\mu\nu}, \qquad
\left\{ \psi^\mu, \psi^\nu \right \} = 0, \qquad \left\{ \bar
\psi^\mu,\bar \psi^\nu \right \} =0. \label{oper-fer}
\end{equation}
There are two forms where the equations in 1-D are satisfied: in the literature we find  either the matrix representation or the differential operator scheme. However for more than one dimensions, there exist many applications to cosmological models, where the differential representation for the Grassmann variables is widely applied \cite{sm,Tkach,s,so,socorro}. There are few works in more dimensions in the first scheme \cite{filho}, we present in this work the main ideas to built the 2D case, where the supercharges operators become  $\rm 4 \times 4$ matrices.
\end{itemize}

\section{Factorization method in 1-Dimension: matrix approach}
We begin by introducing the main ideas for the 1-Dimensional quantum harmonic oscillator . The corresponding hamiltonian is written in operator form as
\begin{equation}
\rm \hat H_B=\frac{1}{2}\hat p^2+\frac{1}{2}\omega_B^2 \hat q^2\label{hamiltonbosonico}
\end{equation}
where $\rm \hat q$ is the generalized coordinate, and $\rm  \hat p$ is the associated momentum,  the canonical commutation relation between this quantities being $\rm [\hat q,\hat p]=i$ . We introduce two new operators, known as the creation and annihilation operators $\rm \hat a^+ ,  \hat a^-$ respectively, defined as
\begin{equation}
 \hat a^-= \frac{1}{\sqrt{2\omega_B}}(\hat p-i\omega_B \hat q), \qquad
\hat a^+=\frac{1}{\sqrt{2\omega_B}}(\hat p+i\omega_B \hat q), \label{creationanihilationbos}
\end{equation}

This hamiltonian can be written in terms of the anti-commutation relation between these operators as
\begin{equation}
 \hat H_B=\frac{\omega_B}{2}\{\hat a^+,\hat a^-\},  \label{Hambos}
\end{equation}
the symmetric nature of $\rm \hat H_B$ under the interchange of $ \hat a^-$ and  $ \hat a^+$ suggests that these operators satisfy  Bose-Einstein statistics, and it is therefore called bosonic.

Now, we build the operators  $ \hat b^-$ and  $ \hat b^+$ that obey  similar rules to operators $ \hat a^-, \hat a^+$ changing $\rm [\, ,\, ] \leftrightarrows \{\, , \, \}$, that is
\begin{equation}
\{\hat b^-,\hat b^+\}=1;\hspace{1 cm} \{\hat b^-,\hat b^-\}=\{\hat b^+,\hat b^+\}=0, \label{fer-ope}
\end{equation}
and in analogy to  (\ref{Hambos}), we define the corresponding new hamiltonian as
\begin{equation}
\rm \hat H_F=\frac{\omega_F}{2}[\hat b^+, \hat b^-], \label{hamfer}
\end{equation}
The antisymmetric nature of $\rm \hat H_F$ under the interchange of $ \hat b^-$ and  $ \hat b^+$ suggests that these operators satisfy the Fermi-Dirac statistics, and it is called fermionic.

These operators $ \hat b^-$ and  $ \hat b^+$ admit a matrix representations in terms of Pauli matrices, that satisfy all rules defined above, that is
\begin{equation}
 \hat b^- = \sigma_- , \qquad  \hat b^+ = \sigma_+, \qquad \sigma_\pm=\frac{1}{2}(\sigma_1\pm i\sigma_2) \label{creationanihilationfer}
\end{equation}
with $\rm [\sigma_+,\sigma_-]=\sigma_3$, $
\sigma_-=
\begin{pmatrix}
0&0\\
1&0
\end{pmatrix},
\quad
\sigma_+=
\begin{pmatrix}
0&1\\
0&0
\end{pmatrix}, \quad \sigma_1 = \begin{pmatrix} 0 & 1\\1&0 \end{pmatrix}, \quad \sigma_2 = \begin{pmatrix} 0&-i\\i&0 \end{pmatrix},
\quad \sigma_3 = \begin{pmatrix} 1&0\\0&-1 \end{pmatrix}. $

Now, consider both hamiltonians as a composite system, that is, we consider the superposition of two oscillators, one being bosonic and one fermionic,  with energy $\rm E_T=E_B+ E_F$
\begin{equation}
\rm E_T=\omega_B(n_B+\frac{1}{2})+\omega_F(n_F-\frac{1}{2})=\omega_B n_B + \omega_F n_F +\frac{1}{2}(\omega_B-\omega_F). \label{enertotiso}
\end{equation}
When we demand that both frequencies are the same, $\omega_B=\omega_F=\omega$, we introduce a new symmetry, called supersymmetry (SUSY),
we can see that the simultaneous creation of a quantum fermion $(n_F\rightarrow n_F+1)$, causes  the destruction of quantum boson
$(n_B\rightarrow n_B-1)$ and viceversa, in the sense that the total energy is unaltered. The ground energy state is exact and no degenerate. The degeneration appears from n=1, where it is double degenerate.

In this way, we have the super-hamiltonian $\rm \hat H_{susy}$, written as
\begin{equation}
\hat H_{susy}=\frac{\omega}{2} \{\hat a^+, \hat a^- \}+ \frac{\omega}{2}[\hat b^+, \hat b^-]=\frac{\omega}{2} \{\hat a^+, \hat a^- \}I+\frac{\omega}{2}\sigma_3=\omega \left(
\begin{tabular}{ll}
$ \hat a^- \hat a^+$ & 0\\ 0 & $\hat a^+ \hat a^-$ \end{tabular} \right)=\begin{pmatrix} \hat H_-&0\\0& \hat H_+ \end{pmatrix}, \label{superhammatrix}
\end{equation}
where I is a $\rm 2\times 2$ unit matrix, and  where the two components of $\rm \hat H_{susy}$ in (\ref{superhammatrix}) can be written  independently  as
\begin{eqnarray}
 \hat H_+=\frac{1}{2}\hat p^2+\frac{1}{2}(\omega^2q^2- \omega)\equiv \omega \hat a^+ \hat a^- \label{ham+}\\
\hat H_-=\frac{1}{2} \hat p^2+\frac{1}{2}(\omega^2q^2+\omega)\equiv \omega \hat a^- \hat a^+\label{ham-}.
\end{eqnarray}
From  equations (\ref{ham+}) and (\ref{ham-}), we can see that  $\rm \hat H_+$ and  $\rm \hat H_-$ are the same representation of one hamiltonian with a constant shifting $\omega$ in the energy spectrum.

The question is, what are the generators for this SUSY hamiltonian? The answer is, considering that the degeneration is the result of the simultaneous destruction (creation) of quantum boson and the creation (destruction) of quantum fermion, that the corresponding generators for this symmetry must be written as
$ \hat a^- \hat b^+$ (or $\hat a^+ \hat b^-$). therefore we introduce the following generators, called  supercharges $\rm \hat Q^-$ and
$\rm \hat Q^+$ defined as
\begin{equation}
 \hat Q^-=\sqrt{2\omega} \hat  a^- \hat b^+=\sqrt{2\omega}\begin{pmatrix} 0 & \hat a^-\\0&0 \end{pmatrix} , \qquad  \hat Q^+=\sqrt{2\omega }\hat a^+ \hat b^-=\sqrt{2\omega}\begin{pmatrix} 0 & 0\\\hat a^+&0 \end{pmatrix} , \label{superchar}
\end{equation}
implying that
\begin{equation}
\rm \hat H_{susy}=\frac{1}{2}\{\hat Q^+,\hat Q^-\} \label{superham}
\end{equation}
and satisfying the following relations
\begin{equation}
\rm \{\hat Q^-, \hat Q^-\}=\{\hat Q^+, \hat Q^+\}=0;\hspace{0.5 cm} [\hat Q^-,\hat H_{susy}]=[\hat Q^+, \hat H_{susy}]=0. \label{superchargconmut}
\end{equation}

We can generalize this procedure for a certain function W(q), and
at this point we can define two new operators $\rm \hat A^-$ and $\rm \hat A^+$ with a property similar  to (\ref{creationanihilationbos}),
\begin{equation}
\rm \hat A^-=\frac{1}{\sqrt{2\omega}}(\hat p-i\omega W(q)), \qquad
\hat A^+=\frac{1}{\sqrt{2\omega}}(\hat p+ i \omega W(q)), \label{new}
\end{equation}

In order to obtain the general solutions, we can use an arbitrary potential in equation (\ref{hamiltonbosonico}), that is
\begin{equation}
\rm \hat H_B=\frac{1}{2}\hat p^2+ V(q),\label{arbitrary-potential0}
\end{equation}
the hamiltonians $\rm \hat H^+$ and $\rm \hat H^-$ determine two new potentials,
\begin{eqnarray}
\rm \hat H_+=\frac{1}{2}\hat p^2+V_+=\frac{1}{2}\hat p^2+\frac{1}{2}\left(W^2- \frac{dW}{dq}\right) \label{ham+}\\
\rm \hat H_-=\frac{1}{2}\hat p^2 + V_-=\frac{1}{2} \hat p^2+\frac{1}{2}\left(W^2+ \frac{dW}{dq}\right),\label{ham-}
\end{eqnarray}
where the potential term V$_+$(q) is related to the superpotential function W(q) via the Ricatti equation
\begin{equation}
\rm V_+=\frac{1}{2}\left(W^2- \frac{dW}{dq}\right), \label{ricatti}
\end{equation}
(modulo constant $\epsilon$, which is related to some energy eigenvalue)
and $\rm V_-=\frac{1}{2}\left(W^2+ \frac{dW}{dq}\right)=V_+ + \frac{dW}{dq}$,
with the same spectrum, except for the ground state, which is related to the energy potential $\rm V_+$.

In a general way, let us now find the general form of the function W. The quantum equation (\ref{arbitrary-potential0}) applied to stationary wave function $\rm u_i$ becomes
\begin{equation}
\rm - \frac{1}{2} \frac{d^2 u_i}{dq^2} + V(q)u_i=E_i u_i, \label{solution}
\end{equation}
where $\rm E_i$ are the energy eigenvalues. Considering the transformation $\rm W(q)=-\frac{d ln[u_i(q)]}{dq}$ and introducing it into (\ref{ham+}), we have that
$$\rm V(q)-E_i=\frac{1}{2}\left(W^2-\frac{dW}{dq}\right)=\left( \frac{1}{2u_i} \frac{du_i}{dq}\right)^2- \frac{\left( \frac{du_i}{dq}\right)^2-u_i \frac{d^2 u_i}{dq^2}}{2u_i^2}= \frac{1}{2u_i}\frac{d^2 u_i}{dq^2}$$
then, this equation is the same as the original one, eq.(\ref{solution}), that is, W is related to a initial solution of the bosonic hamiltonian.
What happens to the iso-potential $\rm V_-(q)=\frac{1}{2}\left(W^2+\frac{dW}{dq}\right)$? Considering  that
$$\rm 2V_-=W^2+\frac{dW}{dq}\equiv \hat W^2+\frac{d \hat W}{dq}=2 \hat V_-,$$
the question is, what is $\rm \hat W$ if we know the function W? Finding it we can  build a family of potentials $\rm \hat V_-$ depending on a free parameter $\lambda$, the supersymmetric parameter that, to some extent, plays the role of  internal time. Following the procedure $\rm \hat W= W+\frac{1}{y(q)}$, where the function y(q) satisfy the linear differential equation $\rm \frac{dy}{dq}-2 Wy=1$, the solution implies
\begin{equation}
\rm y(q)=\frac{\lambda + \int u_i^2 dq}{u_i^2},  \qquad \to \qquad  \hat W=W+\frac{u_i^2}{\lambda + \int u_i^2 dq}. \label{y-b}
\end{equation}
The family of potentials $\rm \hat V_+$ can be built now as
\begin{equation}
\rm \hat V_+-E_i=\frac{1}{2}\left(\hat W^2-\frac{d \hat W}{dq}\right)= V_- + \frac{d \hat W}{dq} \ .
\label{family-pot}
\end{equation}
Finally
\begin{equation}
\rm \hat u=g(\lambda)\frac{u_i}{\lambda+ \int u_i^2 dq}
\end{equation} is the isospectral solution of the Schr\"odinger like equation for the new family potential (\ref{family-pot}), with the condition $\rm g(\lambda)=\sqrt{\lambda (\lambda +1)}$, which in the limit
$$\rm  \lambda \to \pm \infty, \qquad g(\lambda)=\lambda, \qquad \hat u_i\to u_i.$$
This $\lambda$ parameter is included not for factorization reasons; in particular, in quantum cosmology the wave functions are still nonnormalizable, and $\lambda$ is used as a decoherence parameter embodying a sort of quantum cosmological dissipation (or damping) distance.

\subsection{Two dimensional case.}
We use Witten's idea \cite{witten}, to find the supersymmetric
supercharges operators $\rm Q^-$ and $\rm Q^+$ that generate the
superHamiltonian $\rm H_{susy}$.  Using equations (\ref{superchar}), (\ref{superham}) and (\ref{superchargconmut}), we can generalize the one-dimensional factorization scheme.
We define the two dimensional Hamiltonian as
\begin{equation}
\rm \hat H_B(x,y)=\frac{1}{2}\hat p_x^2+ \frac{1}{2}\hat p_y^2+ V(x)+V(y),\label{arbitrary-potential}
\end{equation}
where the Schr\"odinger like equation can be
obtained as the bosonic sector of this super-Hamiltonian in the
superspace, i.e, when all fermionic fields are set equal to zero
(classical limit).

In two dimensions the supercharges are defined by the tensorial products
\begin{equation}
\rm Q^-=\sqrt{2} d|^-\otimes \sigma_+, \qquad
Q^+ = \sqrt{2}d|^+\otimes \sigma_-\label{supercargas}
\end{equation}
with
\begin{eqnarray}
d|^-=
\begin{pmatrix}
a^-&0\\
0&b^-
\end{pmatrix},
\qquad
d|^+=
\begin{pmatrix}
a^+&0\\
0&b^+
\end{pmatrix},
 \label{realization}
\end{eqnarray}
where $\rm \sigma_\pm$ are the same as in (\ref{creationanihilationfer}).
From equations  (\ref{supercargas}) we have that the supercharges are $\rm 4 \times 4$ matrices
\begin{equation}\rm
\hat Q^+=\sqrt{2}
\begin{bmatrix}
0&0&0&0\\
0&0&0&0\\
a^+&0&0&0\\
0&b^+&0&0
\end{bmatrix}
\qquad
\hat Q^-=\sqrt{2}
\begin{bmatrix}
0&0&a^-&0\\
0&0&0&b^-\\
0&0&0&0\\
0&0&0&0
\end{bmatrix}
\end{equation}
where the super-Hamiltonian, (\ref{superham}), can be written as
\begin{equation}
\rm H_{susy}=
\begin{pmatrix}
a^-a^+&0&0&0\\
0&b^-b^+&0&0\\
0&0&a^+a^-&0\\
0&0&0&b^+b^-
\end{pmatrix}=\begin{pmatrix}
H^{1}_-(x)&0&0&0\\
0&H^{1}_-(y)&0&0\\
0&0&H^{2}_+(x)&0\\
0&0&0&H^{2}_+(y)
\end{pmatrix},
\end{equation}
where
\begin{eqnarray}
a^-=\rm \frac{1}{\sqrt{2}}\left(\frac{d}{dx}+W(x)\right), &\qquad& a^+=\frac{1}{\sqrt{2}}\left(-\frac{d}{dx}+W(x)\right)\\
b^-=\rm \frac{1}{\sqrt{2}}\left(\frac{d}{dx}+Z(y)\right), &\qquad& b^+=\frac{1}{\sqrt{2}}\left(-\frac{d}{dx}+Z(y)\right)
\end{eqnarray}
and $\rm V(x,y)=W(x)+Z(y)$.\\

The Ricatti equation (\ref{ricatti}) is written in 2D as
\begin{equation}
\rm V_+(x,y)=V_{+}1(x)+V_{+2}(y)=\frac{1}{2}\left(W^2-\frac{dW}{dx}\right) + \frac{1}{2}\left( Z^2-\frac{dZ}{dy}\right),
\end{equation}
and, using separation variables, we get
\begin{eqnarray}\rm
V_1(x)-\frac{1}{2}\left(W^2(x)-\frac{dW}{dx}\right)=C_0\\
\rm V_2(y)-\frac{1}{2}\left(Z^2(y)-\frac{dZ}{dy}\right)=-C_0
\end{eqnarray}

In general, we find that each potential  $\rm V_{+i}$ satisfy
\begin{equation}\rm
\frac{1}{2}\frac{d^2}{dx^2}u_i(x)+V_{+i}u_i(x)=E_i u_i(x), \qquad  i=1,2,
\end{equation}
and we can find the iso-potential as $\rm W=-\frac{1}{u_1}\frac{du_1}{dx}$, when $\rm u_1$ is known.

Following the same steps as in the 1D case, we find that the solutions (\ref{y-b}) are the same in this case. So, the general solution for $\rm \hat W$ is  $\rm \hat W=W+\frac{1}{y(x)}$, with $\rm y = u_1^{-2}(x)\left[\lambda_1+\int{u_1^2(x)dx}\right]$.
The general solution for the superpotential $\hat W(x)$ is
\begin{equation}
\rm \hat W=-\frac{1}{u_1}\frac{du_1}{dx}+\frac{u_1^2}{\lambda_1+\int{u_1^2\, dx}}
=W_p+\frac{d}{dx}\left[Ln(\lambda_1+I_1)\right]
\end{equation}
where $\rm W_p=-\frac{1}{u_1}\frac{du_1}{dx}$ and $I_1=\int{u_1^2\,dx}$.\\

In the same manner, we have that
\begin{equation}
\rm \hat Z=-\frac{1}{u_2}\frac{du_2}{dy}+\frac{u_2^2}{\lambda_2+\int{u_2^2 \,dy}}
=Z_p+\frac{d}{dy}\left[Ln(\lambda_2+I_2)\right]
\end{equation}
with $\rm Z_p=-\frac{1}{u_2}\frac{su_2}{dy}$ and $\rm I_2=\int{u_2^2 \,dy}$.

On the other hand, using the Ricatti equation, we can build a generalization for the isopotential, using the new potential $\rm \hat W$, as
\begin{equation}\rm
\hat V_{+1}(x,\lambda_1)=\frac{1}{2}\left(\hat{W}^2-\hat{W}'\right)
=V_+(x)-\frac{2u_1 \frac{du_1}{dx}}{\lambda_1+I_1}+\frac{u_1^4}{(\lambda_1+I_1)^2} \label{iso_x}
\end{equation}

For the other coordinate,   we have
\begin{equation}
\hat V_{+2}(y,\lambda_2)=\frac{1}{2}\left(\hat{Z}^2-\frac{d\hat Z}{dy} \right)
=V_+(y)-\frac{2u_2 \frac{du_2}{dy} }{\lambda_2+I_2}+\frac{u_2^4}{(\lambda_2+I_2)^2}. \label{iso_y}
\end{equation}

The general solutions for $\rm \hat u_i$ depends on the initial solutions to the original Schr\"odinger equations in the variables (x,y), that is, $\rm u_1=u_1(x)$, $\rm u_2=u_2(y)$, being
\begin{equation}
\hat u_1(x,\lambda_1)=C_1(\lambda_1)\frac{ u_1}{\lambda_1+I_1}, \qquad \hat u_2(y,\lambda_2)=C_2(\lambda_2)\frac{ u_2}{\lambda_2+I_2}.
\label{ui-general}
\end{equation}
where the {\it variables $C_i(\lambda_i)$} have the same properties that $\rm g(\lambda)$ obtained in the 1D case.

\subsection{Application to cosmological Taub model}
The Wheeler-DeWitt equation for the cosmological Taub model is given by
\begin{equation}
\frac{\partial^2\Psi}{\partial\alpha^2}-\frac{\partial^2\Psi}{\partial\beta^2}+e^{4\alpha}V(\beta)\Psi=0 \label{taub}
\end{equation}
where $\rm V(\beta)=\frac{1}{3}\left(e^{-8\beta}-4e^{-2\beta}\right)$. This equations can be separated using $\rm x_1=4\alpha-8\beta$ and $\rm x_2=4\alpha-2\beta$, rendering
\begin{equation}
\rm -\frac{\partial^2f_1(x_1)}{\partial x_1^2}+\frac{1}{144}e^{x_1}f_1(x_1)=  \frac{\omega^2}{4}f_1(x_1), \qquad
 -\frac{\partial^2f_2(x_2)}{\partial x_2^2}+\frac{1}{9}e^{x_2}f_2(x_2)= \omega^2f_2(x_2),
\end{equation}
where the parameter $\omega$ is the separation constant.
These equations possess the solutions
\begin{equation}
\rm f_1 = K_{i\omega}\left(\frac{1}{6}e^{\frac{x_1}{2}}\right) \label{f1},
\qquad f_2=  L_{2i\omega}\left(\frac{2}{3}e^{\frac{x_2}{2}}\right)+K_{2i\omega}\left(\frac{2}{3}e^{\frac{x_2}{2}}\right)
\end{equation}
where K (or I) is the modified Bessel function of imaginary order, and
the functions L is define as
$$L_{2i\omega}=\frac{\pi i}{2\sinh(2\omega\pi)}\left(I_{2i\omega}+I_{-2i\omega}\right) \ .$$
Using equations (\ref{iso_x}) and (\ref{iso_y}) we obtain the isopotential for this model
\begin{equation}
\rm \hat V(x_1)=V_+(x_1)-\frac{2K_{i\omega}K_{i\omega}'}{\lambda_1+I_1}+\frac{K_{i\omega}^4}{\left(\lambda_1+I_1\right)^2}, \qquad
\hat V(x_2)=V_+(x_2)-\frac{2\left(L_{2i\omega}+K_{2i\omega}\right)\left(L_{2i\omega}+K_{2i\omega}\right)'}{\lambda_2+I_2}
+\frac{\left(L_{2i\omega}+K_{2i\omega}\right)^4}{\left(\lambda_2-I_2\right)^2}
\end{equation}
Using (\ref{ui-general}) we can obtain general solutions for the functions $\rm f_1$  and  $\rm f_2$ in the following way
\begin{equation}
\hat f_1=\frac{C_1K_{i\omega}\left(\frac{1}{6}e^{\frac{x_1}{2}}\right)}{\lambda_1+I_1}, \qquad
\hat f_2=\frac{C_2\left[L_{2i\omega}\left(\frac{2}{3}e^{\frac{x_2}{2}}\right)
+K_{2i\omega}\left(\frac{2}{3}e^{\frac{x_2}{2}}\right)\right]}{\lambda_2+I_2}
\end{equation}


\section{ Differential approach: Grassmann variables}
The supersymmetric scheme has the particularity of being very restrictive, because there are many constraint equations applied to the wave function. So, in this work and in others, we
found that there exist a tendency for supersymmetric vacua to remain close to their semiclassical limits, because the exact solutions found are also the lowest-order WKB like approximations, and do not correspond to the full quantum solutions found previously for particular
models.\cite{sm,Tkach,s,so,socorro}

 Mantaining  the structure of the equations (\ref{superchar}), (\ref{superham}), (\ref{superchargconmut}) and (\ref{new}), taking the differential representation for the fermionic operator $\hat b \leftrightarrow \psi^\mu$ for convenience in the calculations, and changing the function $\rm W \to \frac{\partial S}{\partial q^\mu}$, the  supercharges for the n-dimensional case  read as
\begin{equation}
\rm \hat Q^- = \psi^\mu \left[P_\mu + i\frac{\partial
S}{\partial q^\mu} \right], \qquad \rm \hat Q^+ = \bar \psi^\nu
\left[P_\nu - i\frac{\partial S}{\partial q^\nu}
\right], \label{supercharge2}
\end{equation}
where $\rm S$ is known as the super-potential functions which are related to the physical potential under consideration, when the hamiltonian density is written as the Hamilton-Jacobi equation,
 and the following algebra for the variables  $\psi^\mu$ and $\bar \psi^\nu$, (similar to equation (\ref{fer-ope}))
\begin{equation}\rm
\left\{ \psi^\mu ,\bar \psi^\nu \right \} = \eta^{\mu\nu}, \qquad
\left\{ \psi^\mu, \psi^\nu \right \} = 0, \qquad \left\{ \bar
\psi^\mu,\bar \psi^\nu \right \} =0. \label{oper-fer}
\end{equation}
these rules are satisfied when we use a differential representation for these $\psi^\mu , \bar \psi^\nu$ variables in terms of the Grassmann numbers, as
\begin{equation}
\rm \psi^\mu=\eta^{\mu\nu} \frac{\partial}{\partial \theta^\nu}, \qquad\qquad \bar \psi^\nu=\theta^\nu,
\end{equation}
where $\eta^{\mu\nu}$ is a diagonal constant matrix, its dimensions depending on the independent bosonic variables that appear in the bosonic hamiltonian.  Now the superhamiltonian is written as
\begin{equation}
H_S=\frac{1}{2}\{\hat Q^+,\hat Q^-\}={\cal H}_0 + \frac{\hbar}{2}\frac{\partial^2 S}{\partial q^\mu \partial q^\nu} \left[ \bar \psi^\mu, \psi^\nu \right], \label{ss}
\end{equation}
where $\rm {\cal H}_0 =\Box + U(q^\mu)$ is the quantum version of the classical bosonic hamiltonian, $\Box$ is the d'Alambertian in three dimension when we have three bosonic independent coordinates, and $\rm U(q^\mu)$ is the  potential energy in consideration.

The superspace for three dimensional model becomes $(q_1,q_2,q_3,\theta^0, \theta^1, \theta^2)$, where the variables $\theta^i$ are the coordinate in the fermionic space, as the Grassmann numbers, which have the property of $\theta^i \theta^j=-\theta^j \theta^i$, and the wavefunction has the representation
\begin{eqnarray}
\rm \Psi&=& \rm {\cal A}_+ + {\cal B}_0 \theta^0, \qquad 1 \,dimension \label{1d}\\
\rm \Psi&=& \rm {\cal A}_+ + {\cal B}_0 \theta^0 + {\cal B}_1 \theta^1 + {\cal A}_- \theta^0 \theta^1, \qquad 2\, dimensions \label{2d}\\
\rm \Psi&=& \rm {\cal A}_+ + {\cal B}_\nu \theta^\nu +\frac{1}{2} \epsilon_{\mu \nu \lambda} {\cal C}^\lambda \theta^\mu \theta^\nu + {\cal A}_- \theta^0 \theta^1 \theta^2, \qquad 3\, dimensions \label{3d}
\end{eqnarray}
where the indices $\rm \mu, \nu, \lambda$ values are 0,1 and 2, and $\rm {\cal A}_\pm, {\cal B}_\nu$ and $\rm {\cal C}^\lambda$
are bosonic functions which depend on the bosonic coordinates $\rm q^\mu$ and not on the Grassmann numbers.  Here, the wavefunction representation structure is set in terms of $2^n$ components, for $n$ independent bosonic coordinates, with half of the terms coming from the  bosonic (fermionic) contribution into the wavefunction.

It is well known that the physical states are determined by the applications of the supercharges $\hat Q^-$ and $\hat Q^+$ on the wavefunctions, that is
\begin{equation}
\rm  \hat Q^- \Psi=0, \qquad \hat Q^+ \Psi=0, \label{states}
\end{equation}
where we use the usual representation for the momentum $\rm P_\mu=-i\hbar \frac{\partial}{\partial q^\mu}$.
Considering the 2D case, the last second equation gives
\begin{eqnarray}
\theta^0 &:& \left [ \frac{\partial A_+}{\partial q^0}-A_+\frac{\partial S}{\partial q^0}\right] =0,\label{tetabar0}\\
\theta^1 &:& \left [ \frac{\partial A_+}{\partial q^1}-A_+\frac{\partial S}{\partial q^1}\right] =0,\label{tetabar1} \\
\theta^0\theta^1 &:& \left [ \frac{\partial B_1}{\partial q^0}-B_1\frac{\partial S}{\partial q^0}\right]-
\left [ \frac{\partial B_0}{\partial q^1}-B_0\frac{\partial S}{\partial q^1}\right] =0,\label{tetabar01}
\end{eqnarray}
from (\ref{tetabar0}) and (\ref{tetabar1}) we obtain the relation
 $\frac{\partial A_+}{\partial q^\mu}-A_+\frac{\partial S}{\partial q^\mu}=0$ with the solution
$\rm A_+=a_+e^S.$

On the other hand, the first equation in (\ref{states}) gives
\begin{eqnarray}
\theta^0 &:& \left [ \frac{\partial A_-}{\partial q^1}+A_-\frac{\partial S}{\partial q^1}\right] =0,\label{teta0}\\
\theta^1 &:& \left [ \frac{\partial A_-}{\partial q^0}+A_-\frac{\partial S}{\partial q^0}\right] =0,\label{teta1}\\
free\, term &:& -\left [ \frac{\partial B_0}{\partial q^0}+B_0\frac{\partial S}{\partial q^0}\right]+
\left [ \frac{\partial B_1}{\partial q^1}+B_1\frac{\partial S}{\partial q^1}\right] =0,\label{tetalibre}
\end{eqnarray}
the free term equation is written as $\rm \eta^{\mu\nu}(\partial_\mu B_\nu+B_\nu\partial_\mu S)=0$, and
taking the ansatz $B_\mu=e^{-S}\partial_\nu f_+(q^\mu),$
the equation (\ref{tetabar01}) is fulfilled, so we obtain for the free term,
\begin{equation}
\rm \square f_+ + 2\eta^{\mu\nu} \nabla_\mu S \nabla_\nu f_+=0, \label{master+}
\end{equation}
with the solution to  $f_+=h(q^1-q^2)$, with h an arbitrary function depending of its argument. However, this function f must depend on the potential under consideration.

Also, equations (\ref{teta0}) and (\ref{teta1}) are written as
\begin{equation}
\rm \frac{\partial A_-}{\partial q^\mu}+ A_- \frac{\partial S}{\partial q^\mu}=0, \qquad
\frac{1}{A_-} \frac{\partial A_-}{\partial q^\mu}=- \frac{\partial S}{\partial q^\mu} \qquad \rightarrow
\qquad \frac{\partial Ln A_-}{\partial q^\mu }=-\frac{\partial S}{\partial q^\mu}
\end{equation}
whose solution is $\rm A_-=a_- e^{-S}$.In this way, all functions entering the wavefunction are
$$\rm A_\pm =a_\pm e^{\pm S}, \qquad
B_0 = e^{-S}\partial_0 (f_+),  \qquad
B_1 = e^{-S}\partial_1 (f_+).$$


\subsection{The unnormalized probability density}
To obtain the wavefunction probability density $\rm |\Psi|^2$ in this supersymmetric fashion, we need first to integrate over the Grassmann variables $\theta^i$. This procedure is well known,\cite{faddeev} and here we present the main ideas. Let $\rm \Psi_1$
and $\rm \Psi_2$ be two functions that depend on Grassmann numbers, the product $<\Psi_1,\Psi_2>$ is defined as
$$<\Psi_1,\Psi_2>=\int (\Psi_1(\theta^*))^* \Psi_2(\theta^*)\, e^{-\sum_i \theta^*_i \theta_i } \Pi_i d\theta^*_i d \theta_i, \qquad (C \theta_i \cdots \theta_r)^*=\theta^*_r \cdots \theta^*_i C^*,$$
and the integral over the Grassmann numbers is $\int \theta^*_i \theta_i \cdots \theta_m \theta_m^* d\theta^*_m  d\theta_m \cdots d\theta^*_i d\theta_i =1$.

In 2D, the main contributions to the term $e^{-\sum_i \theta^*_i \theta_i }$ come from
$$e^{-\sum_i \theta^*_i \theta_i}=e^{\sum_i \theta_i \theta^*_i}=1+ \theta^0 \theta^{*0} + \theta^1 \theta^{*1} +
\theta^0 \theta^{*0}  \theta^1 \theta^{*1}$$
and using that $\rm \int \theta d\theta=1$, and $\int d\theta=0$, which act as a filter, we obtain that
$$\rm |\Psi|^2={\cal A^*_+} {\cal A_+} +{\cal B}^*_0 {\cal B}_0 +{\cal B}^*_1 {\cal B}_1 + {\cal A^*_-} {\cal A_-}.$$
By demanding that $\rm |\Psi|^2$ does not diverge when $\rm |q^0|, |q^1|\to \infty$, only the contribution with the exponential $\rm e^{-2S}$ will remain.




\section{Beyond SUSY factorization\label{susy2par1}}

Although most of the SUSY partners of 1D Schr\"odinger problems have been found,\cite{cooper} there are still some unveiled aspects of the factorization procedure.
We have shown this for the simple harmonic oscillator in previous works,\cite{ranferi,rafael} and will procede here in the same way for the problem of the modified P\"oschl-Teller potential. The factorization operators depend on two supersymmetric type parameters, which when the operator product is inverted allow us to define a new SL operator, which includes the original QM problem.


The Hamiltonian of a particle in a modified P\"oschl-Teller potential is \cite{rosen,cooper}
\begin{equation}
 H_{m+1} \, \Psi =
 \left(-\frac{\hbar^2}{2\mu}\frac{d^2}{dx^2}-\frac{\alpha^2m(m+1)}{\cosh^2\!\alpha x}\right)\Psi=E\,\Psi \ ,
 \label{eq1}
\end{equation}
where $\alpha>0$, and the integer $m$ is greater than 0.  To shorten the algebraic equations we shall set $\frac{\hbar^2}{2\mu}=1$.

The eigenvalue problem may be solved using the Infeld \& Hull's (IH)
factorizations,\cite{infeld}
\begin{subequations}\label{fih}
\begin{align}
 \label{fih1}
 A^+_{m+1} A^-_{m+1} \, \psi^m_{m-n} &=
 \left(H_{m+1}+\epsilon_{m+1}\right) \psi^m_{m-n} , \\
 \label{fih2}
 A^-_m A^+_m \, \psi^m_{m-n} &=
 \left(H_{m+1}+\epsilon_{m}\right) \psi^m_{m-n} ,
\end{align}
\end{subequations}
where the IH raising/lowering operators are given by
\begin{equation}\label{ampih}
 A^\mp_{m}=
k(x,m)\mp\frac{d}{dx}
\:.
\end{equation}
where 
$k(x,m)=\alpha m\, \tanh\alpha x$; also $\epsilon_m=\alpha^2 m^2$, and $n$ is the eigenvalue index,
\begin{equation}\label{esubn}
 \Psi_n=\psi_{m-n}^m, \hspace{5mm} E_{n}=-\epsilon_{m-n}=-\alpha^2(m-n)^2,    \hspace{1cm} n=0,1,2...<m .
\end{equation}
Beginning with the zeroth order eigenfunctions
The eigenfunctions can be found by successive applications of the raising operator,
which only increases the value of the upper index. That is,
\begin{equation}\label{psi0}
 \psi_\ell^\ell(x)=\sqrt{\frac{\alpha\Gamma(\ell+\frac{1}{2})}{\sqrt{\pi}\Gamma(\ell)}}\cosh^{-\ell}\alpha x .
\end{equation}
we repeatedly apply the creation operator $A^-_{s+1} \, \psi^s_\ell = \psi^{s+1}_{\ell}$.
Note that from (\ref{fih}), $A^-_m A^+_m$ and $A^+_m A^-_m$ give different Hamiltonian operators.


\subsection{Two parameter factorization of the P\"oschl-Teller Hamiltonian}

Following our previous work,\cite{ranferi,rafael} we define two non-mutually adjoint first order operators,
\begin{equation}\label{bmbp}
 B_m=\eta_m^{-1}\frac{d}{dx}+\beta_m,   \hspace{2cm}
 B^*_m=-\eta_m\frac{d}{dx}+\beta_m,
\end{equation}
where $\beta_m$ and $\eta_m$ are functions of $x$, and we require that
$B_{m+1}B^*_{m+1}=H_{m+1}+\epsilon_{m+1}$.  Then $\beta_{m+1}$ and $\eta_{m+1}$ are the solutions of
\begin{equation}\label{coup1p}
 -\frac{\eta'}{\eta} + \frac{\beta}{\eta} - \beta\eta = 0,
\hspace*{1.5cm}
 \frac{\beta'}{\eta}+\beta^2 =
 - \frac{\alpha^2 m(m+1)}{\cosh^2\alpha x} + \epsilon\,.
\end{equation}
By multiplying the first equation by $\beta/\eta$ and adding, we have that
\begin{equation}\label{ricmp}
 \left( \frac{\beta_{m+1}}{\eta_{m+1}} \right)'+
 \left( \frac{\beta_{m+1}}{\eta_{m+1}} \right)^2=
 -\frac{\alpha^2 m(m+1)}{\cosh^2\alpha x} + \epsilon_{m+1}\,.
\end{equation}
This Ricatti equation was found in \cite{rosas}, it has the solution $\beta/\eta=D\, \tanh\alpha x$, with $\epsilon=D^2$, and two possible values for $D$, $D = \alpha(m+1) \, , -\alpha m $.
If we simply set $\eta_m \to 1$, we recover the factorization (\ref{fih1}).

The constant $\epsilon$ is usually related  to the lowest energy eigenvalue, but here the two different values come from the index asymmetry in the factorizations (\ref{fih}).
Following Ref.\cite{rosas}, we solve for $D=\alpha (m+1)$.

The general solution to the pair of coupled equations (\ref{coup1p}) is
\begin{equation}\label{eta2p}
 \eta_{m+1}(x)=\left[
  1+\frac{\gamma_2 \sech^{2(m+1)} \alpha x}{
  \left( 1 + \gamma_1 \int_0^x \sech^{2(m+1)}\alpha y\, dy \right)^2}
 \right]^{-1/2},
\end{equation}
and
\begin{equation}\label{beta2p}
\beta_{m+1}(x) =
 \left[\alpha(m+1) \, \tanh\alpha x +
 \frac{\gamma_1 \sech^{2(m+1)}\alpha x}
 {1 + \gamma_1\int_0^x \sech^{2(m+1)}\alpha y\ dy }
\right]
\times \eta_{m+1}(x) \, .
\end{equation}
where $\gamma_1$ has to satisfy
$ |\gamma_1|<2\alpha\,\Gamma(m+3/2) / \left( \sqrt{\pi}\,\Gamma(m+1)\right)$.
The corresponding condition on $\gamma_2$ involves trascendental functions, but one may use
$\gamma_2>-1+\gamma_1^2$ determine the $(\gamma_1,\gamma_2)$ parameter space.
When $\gamma_1=\gamma_2=0$ we recover the original IH raising/lowering operators.


\subsection{Reversing the operator product: new Sturm-Liouville operator}

Now we invert the first order operators' product, keeping in mind eq.(\ref{fih2}),
\begin{eqnarray}\label{bpbm}
 B^*_m B_m=-\frac{d^2}{dx^2}+2 \frac{\eta_m'}{\eta_m} \frac{d~}{dx}+
 \left( V_0+\epsilon_m-\eta_m\beta_m'-\frac{\beta_m'}{\eta_m} \right) \, .
\end{eqnarray}
Then we can define a new Sturm-Liouville (SL) eigenvalue problem
$ \L \Phi_n + \omega(x) E_n \Phi_n=0 $,
where
\begin{equation}\label{opsl2p}
 \L=\frac{d~}{dx}\left[ \eta_{m}^{-2}\frac{d~}{dx}\right]
 +\left(\epsilon_{m}-\beta_{m}^2\right)
 \left(1+\eta_{m}^{-2}\right)
 -\alpha^2m(m+1) \sech^2(\alpha x)
\end{equation}
\begin{equation}\label{phinew}
 \Phi_n=\phi_{m-n}^m\equiv B^*_m\, \psi_{m-n}^{m-1} \, ,
\end{equation}
with the weight function $\omega(x)=\eta_m^{-2}(x)$.

This new SL operator is isospectral to the original PT problem.  The zeroth-order eigenfunction is easily found by solving $B\phi_0= \left[ \frac{d~}{dx} +\beta_m \eta_m\right]\Phi_0=0 $
which gives
\begin{equation}\label{phi0gen2sol}
 \Phi_0= \eta_m(x)\times \frac{\sech^{m+1}(\alpha x)}
 {1+\gamma_1 \int_0^x \sech^{2(m+1)}(\alpha y)\, dy }
\end{equation}
%


\subsection{Regions in the two-parameter space}

We may recover the original QM problem when $\gamma_1=\gamma_2=0$, the origin of the two-parameter space.  Moreover, the SUSY partner of the PT problem arises when one sets $\gamma_2=0$, moving along the horizontal axis.  In this case,
$\L$ becomes
\begin{equation}\label{opsl2psusy}
 \L=\frac{d^2~}{dx^2}+\alpha^2\lambda(\lambda+1) \sech^2(\alpha x)
 -2 S_1^2(\alpha x) -4 \alpha \lambda \tanh(\alpha x) S_1(\alpha x)
\end{equation}
where $\lambda=m+1$, with
$
 S_1(\alpha x)= \frac{\gamma_1 \sech^{2\lambda}(\alpha x)}
 {1+\gamma_1 \int_0^x \sech^{2\lambda}\alpha y\, dy}
$,
and $\omega(x)=1$.  These in turn define a SUSY PT problem
\begin{equation}\label{susy2}
 \left[-\frac{d^2~}{dx^2}+\widetilde V(x)\right] \Phi_n = E_n \Phi_n(x)
\end{equation}
where the partner SUSY potentials are given by
\begin{equation}\label{Vsusy2}
 \widetilde V=-\alpha^2\lambda(\lambda+1) \sech^2(\alpha x)
 +2 S_1^2(\alpha x) +4 \alpha\lambda\, \tanh(\alpha x) \, S_1(\alpha x)
\end{equation}

The zero-order eigenfunction is defined by $B^-\phi_0=0$, that is
\begin{equation}\label{phi02}
 \phi_0= \frac{\sech^\lambda(\alpha x)}
 {1+\gamma_1 \int_0^x \sech^{2\lambda}(\alpha y)\, dy }
\end{equation}
%


\section{Quasi-exactly solvable potentials}

In exactly solvable problems the whole spectrum is found analytically, but the vast majority of problems have to be solved numerically.
A new possibility arised with the class of QES potentials, where a subset of the spectrum may be found analytically.\cite{Turbiner,Shifman,Ushveridze1}
QES potentials have been studied using the Lie algebraic method \cite{Turbiner}:
Manning,\cite{Qiong} Razavy\cite{Razavy}, and Ushveridze\cite{Ushveridze2} potentials belong to this class (see also \cite{Chennn}).
Theses are double well potentials, which received much attention due to their applications in theoretical and experimental problems. Furthermore, hyperbolic type  potentials are found in many physical applications, like the Rosen-Morse potential,\cite{Oyewumi} Dirac type hyperbolic potentials,\cite{Wei} bidimensional quantum dot,\cite{Xie} Scarf type entangled states,\cite{Downing} etc.
QES potentials classification have been given by Turbiner,\cite{Turbiner} and Ushveridze.\cite{Ushveridze2}

Here we show that the Lie algebraic procedure may impose strict restrictions on the solutions:
we shall construct here analytical solutions for the Razavy type potential
$V(x)=V_0\left( {\rm sinh}^4(x)- k\, {\rm sinh}^2(x) \right)$ based on the polynomial solutions of the related Confluent Heun Equation (CHE) \cite{Ronveaux}, and show that in that case the energy eigenvalues diverge when $k\to -1$, a feature solely of the procedure.  We shall also show that other QES potentials may be found that do not belong to any of the potentials found using the Lie algebraic method.



\subsection{A Razavy type QES potential}

Let us consider Schr\"odinger's problem for the Razavy type potential
$V(x)=V_0\left( {\rm sinh}^4(x)- k\, {\rm sinh}^2(x) \right)$,
\begin{equation} \label{ecuaciondeschrodinger}
\frac{-\hbar^2}{2\mu} \frac{d^2 \psi(x)}{d x^2} +
V_0 \left( \sinh^{4}(\lambda x) -k\, {\rm sinh}^2(\lambda x) \right) \,
\psi(x)=E \, \psi(x)
\end{equation}
For simplicity, we set $\mu=\hbar=\lambda=1$.\cite{Downing,Wen}

Here the potential function is the hyperbolic Razavy potential
$V(x)=\frac{1}{2}\left( \zeta \,{\rm cosh}(2x)-M \right)^2$, with $V_0=2\zeta^2$,
where $M$ energy levels are found if $M$ is a positive integer.\cite{Razavy}
It may also be viewed as the Ushveridze potential
$V(x)=2\xi^2 \,{\rm sinh}^4(x)+2\xi\left[ \xi-2(\gamma+\delta)-2\ell \right] {\rm sinh}^2(x)
+2(\delta-\frac{1}{4})(\delta-\frac{3}{4})\, {\rm csch}^2(x)
-2(\gamma-\frac{1}{4})(\gamma-\frac{3}{4})\, {\rm sech}^2(x)$,
when $\gamma=\frac{1}{4}$ and $\delta=\frac{3}{4}$, or viceversa,\cite{Ushveridze2}
which is QES if $\ell=0,1,2,\cdots$ (with $\delta\ge \frac{1}{4}$).
El-Jaick {\it et al.} showed that it is also QES if $\ell=$half-integer and
$\gamma,\delta=\frac{1}{4},\frac{3}{4}$,\cite{Jaick}.

In the case of the Razavy potential, the solutions obtained by Finkel {\it et al.}, are
\begin{equation}
\psi_{\sigma \eta} \left( x,E_{R} \right) \propto \left( \sinh x \right)^{ \frac{1}{2} \left( 1 - \sigma - \eta \right)} \left( \cosh x \right)^{\frac{1}{2} \left( 1 - \sigma + \eta \right)} e^{- \frac{\zeta}{2} \cosh(2x)} \sum_{j=0}^{n} \frac{\hat{P}_{j}^{\sigma \eta} \left(E_{R} \right)}{ \left( 2j + \frac{\eta - \sigma + 1}{2} \right) !} \cosh^{2j}(x)
\end{equation}
with the parameters $(\sigma,\eta)=(\pm 1,0)$ or $(0,\pm 1)$, the energy eigenvalues being the roots of the polynomials $P_{j+1}^{\sigma \eta}(E_R)$, satisfying the three term recursive relations
\begin{equation}
\hat{P}_{j+1}^{\sigma \eta} = \left( E_{R} - b_{j} \right) \hat{P}_{j}^{\sigma \eta} \left( E_{R} \right) - a_{j} \hat{P}_{j-1}^{\sigma \eta} \left( E_{R} \right), \qquad j \geq 0
\end{equation}
with $E_R=2E$, and
\begin{equation}\label{ajbj}
\begin{matrix}
a_j=16\zeta j(2j-\sigma+\eta)(j-n-1)
\\
b_{j} = -4j \left( j + 1 - \sigma + 2 \zeta \right) + \left( 2n + 1 \right) \left( 2 \left( n - \sigma \right) + 3 \right) + \zeta \left( \zeta - 2 \eta + 4n \right)
\end{matrix}
\end{equation}


\subsection{Symmetric solutions for $\boldsymbol{V(x) = V_0 ~ \mathrm{sinh}^{4}(x)}$}\label{seckeq0}

To find the even solutions to eq.(\ref{ecuaciondeschrodinger}) with $k=0$, let us set
$\beta(x) = \cosh^2(x)$, to get
\begin{equation} \label{ecuacion4}
\beta \left( \beta -1 \right) \frac{d^{2} \psi}{d \beta^{2}} + \left( \beta - \frac{1}{2} \right) \frac{d \psi}{d \beta} + \frac{1}{4} \left[ 2 E - 2 V_0 \beta^{2} + 4 V_0 \beta - 2 V_0 \right] = 0
\end{equation}
and to ensure that $\psi (x)$ vanishes as $x\to\pm\infty$, let
$\psi \left( x \right) = e^{-\frac{\alpha}{2} \beta} f(\beta)$.
Previous works may not include square integrable solutions to the Razavy potential.\cite{no2int2,no2int3,no2int}
By requiring $\alpha^{2} = 2 V_{0}$, we obtain \cite{Yao}
\begin{equation} \label{ecuacion5.1}
\beta \left( \beta - 1\right) \frac{d^{2} f}{d \beta^{2}} + \left[ - \alpha \beta \left( \beta -1 \right) + \left( \beta - \frac{1}{2} \right) \right] \frac{d f}{d \beta} + \left[ \frac{\alpha^{2} \beta}{4} - \frac{\alpha \beta}{2} + \frac{\alpha}{4} + \frac{E}{2} - \frac{\alpha^{2}}{4} \right] f = 0 \ .
\end{equation}
We shall look for rank $N$ polynomial solutions: $f(\beta)$=$f_0$ for $N=0$,
or $f(\beta)$=$f_0 \prod_{i=1}^{N} \left( \beta - \beta_i \right)$ for $N>0$, the $\beta_i$ being the roots of the resulting polynomial in eq.(\ref{ecuacion5.1}).  Sometimes the $N$=$0$ solution is not even considered.\cite{Downing}

The highest power of $\beta$ in eq.(\ref{ecuacion5.1})
fix $\alpha$ to $\alpha = 4 N + 2$. The energy eigenvalues and the roots satisfy
\begin{equation}
E = \frac{1}{2} \left[ \alpha^{2} + \alpha \left( 4 \sum_{i=1}^{N} \beta_{i} - 1 - 4 N \right) - 4 N^{2} \right] \label{ecuacion2.101c1}
\end{equation}
\begin{equation} \label{ecuacion2.102c1}
\sum_{i \neq j}^{N} \frac{2}{\beta_{i} - \beta_{j}} + \frac{-\alpha \beta_{i}^{2} + \left( \alpha + 1 \right) \beta_{i} - \frac{1}{2}}{ \beta_{i}^{2} - \beta_{i}} = 0, \qquad i = 1,2,\ldots, n
\end{equation}
$V_0$ is found to depend on the order of the polynomial,
$V_{0} = 2(2N+1)^2$ for even solutions, and solutions with different $N$ can not be scaled one into the other due to the sinh$^4(x)$ dependence of the potential function.  The highest solution order
is $n=2N$, and we use subindexes $\left\{N,n\right\}$ to label eigenvalues/eigenfunctions.


For $N=0$, $f(\beta)=1$,
we get $V_0 = 2$, $E_{0,0} = 1$, and the (unnormalized) ground state eigenfunction
$\psi_{0,0} \left( x \right) = e^{- \cosh^{2} \left( x \right)}$.
For $N = 2$, $f(\beta)=f_0(\beta-\beta_1)(\beta- \beta_2 )$,
equating to zero the coefficients of the polynomial $P(\beta)$, we get the coupled equations
\begin{equation}\label{eq7}
\begin{matrix}
&\frac{\alpha^{2}}{4} - \frac{5 \alpha}{2} = 0 \\
&3+ \left( \beta_1 + \beta_2 \right) \left( - \frac{\alpha^{2}}{4} + \frac{3\alpha}{2} \right) + \left( - \frac{\alpha^{2}}{4} + \frac{9 \alpha}{2} + \frac{E}{2}\right) = 0 \\
&-3 - \left( \beta_1 + \beta_2 \right) \left( - \frac{\alpha^{2}}{4} + \frac{5 \alpha}{4} + \frac{E}{2} + 1 \right) + \beta_1 \beta_2 \left( \frac{\alpha^{2}}{4} - \frac{\alpha}{2} \right) = 0 \\
&\frac{1}{2} \left( \beta_1 + \beta_2 \right) + \beta_1 \beta_2 \left( -\frac{\alpha^{2}}{4} + \frac{\alpha}{4} + \frac{E}{2} \right) = 0
\end{matrix}
\end{equation}
Solving these, we find that $V_{0} = 50$, and the 3 possible eigenvalues, $E_{2,0} = \ 2.6301$,
$E_{2,2} = 19.0121$, and $E_{2,4} = 43.2490$.
%


\subsection{Antisymmetric solutions}\label{seckeq0b}

In order to find antisymmetric solutions to eq.(\ref{ecuacion5.1}), we set
$f(\beta) = \mathrm{sinh}(x) \, g(\beta)$, to obtain
\begin{align} \label{ecuacion32}
\nonumber \beta \left[ \beta - 1 \right] \frac{d^{2}g}{dx^{2}}  &+  \left[ - \alpha \beta^{2} +  \left( \alpha + 2 \right) \beta - \frac{1}{2} \right] \frac{d g}{dx} \\
&+ \left[ \left( - \alpha + \frac{\alpha^{2}}{4} \right) \beta + \left( - \frac{\alpha^{2}}{4} + \frac{\alpha}{4} + \frac{E}{2} + \frac{1}{4} \right) \right] g = 0
\end{align}
This CHE can be solved in power series:
$g(\beta) = g_{0}$ if $N = 0$, or  $g(\beta)=g_{0}\, \prod_{i=1}^{N} \left( \beta - \beta_i \right)$ for $N > 0$.  Then, $\alpha = 4 (N + 1)$, and
\begin{equation}
E = \frac{1}{2} \left[ \alpha^{2} + \alpha \left( 4 \sum_{i=1}^{N} \beta_{i} - 1 - 4 N \right)
 - 4 N^{2} -4N -1 \right] \label{ecuacion2.101c2}
\end{equation}
Here, $V_{0} = 8(N+1)^2$, and all even and odd solutions have different $V_0$.  The maximum solutions order is $n=2N+1$.
For example, for $N = 3$ we get $\alpha=16$, $V_{0} = 128$, and
\begin{equation}
\begin{matrix}
&\left( \beta_1 + \beta_2 + \beta_3 \right) \left( 3 \alpha - \frac{\alpha^{2}}{4} \right) + \left( - \frac{\alpha^{2}}{4} + \frac{13 \alpha}{4} + \frac{E}{2} - \frac{49}{4}\right) = 0 \\
&\left( \beta_1 + \beta_2 + \beta_3 \right) \left( \frac{\alpha^{2}}{4} - \frac{9 \alpha}{4} - \frac{E}{2} - \frac{25}{4} \right) + \left( \beta_1 \beta_2 + \beta_2 \beta_3 + \beta_3 \beta_1 \right) \left( \frac{\alpha^{2}}{4} - 2 \alpha \right) - \frac{15}{2} = 0 \\
&3 \left( \beta_1 + \beta_2 + \beta_3 \right) + \left( \beta_1 \beta_2 + \beta_2 \beta_3 + \beta_3 \beta_1 \right) \left( -\frac{\alpha^{2}}{4} + \frac{5 \alpha}{4} + \frac{9}{4} + \frac{E}{2} \right) + \beta_1 \beta_2 \beta_3 \left( - \frac{\alpha^{2}}{4} + \alpha \right) = 0 \\
&- \frac{1}{2} \left( \beta_1 \beta_2 + \beta_2 \beta_3 + \beta_3 \beta_1 \right) - \beta_1 \beta_2 \beta_3 \left( \frac{\alpha^{2}}{4} - \frac{\alpha}{4} - \frac{E}{2} - \frac{1}{4} \right) = 0
\end{matrix}
\end{equation}
We find four eigenvalues,
$E_{3,1} = 12.8152$,
$E_{3,3} = 40.4568$,
$E_{3,5} = 75.7246$, and
$E_{3,7} = 117.003$.
%



\section{The potential function
$\boldsymbol{V(x) = V_0 \left( \mathrm{sinh}^{4}(x) - k \ \mathrm{sinh}^{2}(x)\right) }$}\label{seckgt0}

Now we apply our analysis to the problem with the
$V(x) = V_0 \left( \mathrm{sinh}^{4}(x) - k \ \mathrm{sinh}^{2}(x)\right)$, which is a symmetric double well if $k>0$.
To find even solutions we set again $\beta(x) = \cosh^{2} (x)$ and
$\psi (\beta) = e^{- \frac{\alpha}{2} \beta} f (\beta)$, with $\alpha^2=2V_0$,
\begin{align}
\label{ecuacion5.5c5}
\nonumber \beta \left( \beta - 1 \right) \frac{d^{2} f}{d \beta^{2}} + &\left[ - \alpha \beta \left( \beta - 1 \right) + \left( \beta - \frac{1}{2} \right) \right] \frac{d f}{d \beta} \\
&+ \left[ \frac{\alpha^{2} \beta}{4} \left(1 + k \right) - \frac{\alpha \beta}{2} + \frac{\alpha}{4} + \frac{E}{2} - \frac{\alpha^{2}}{4} \left( 1 + k\right)\right] f = 0 \ .
\end{align}

We now find that $V_0=\frac{2(2N+1)^2}{1+k}$, $k$ varying freely. For example, if $N=0$,
$E_{0,0}=1/(1+k)$, and no negative energy eigenvalues may exist.
For $N=1$ the two energy eigenvalues found are
\begin{equation} \label{Ek-n1}
E = \frac{9 - \left( 1+ k \right) \pm \sqrt{ \left( 1+ k \right)^{2} + 36 }}{1+k}
\end{equation}
meaning that for $k>3/2$ we will have negative eigenvalues.  Note that for $N>0$ it is always possible to find a zero-energy groundstate, a feature that may have cosmological implications.\cite{socorro}

\begin{figure}
\centering
{\includegraphics[width=0.45\textwidth]{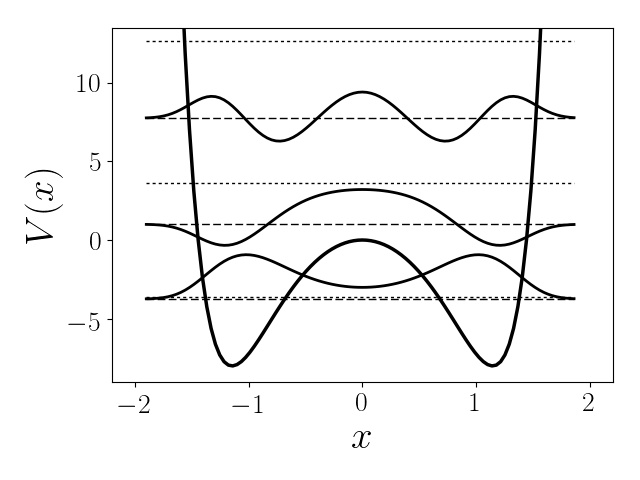}}
\hspace{7mm}
{\includegraphics[width=0.45\textwidth]{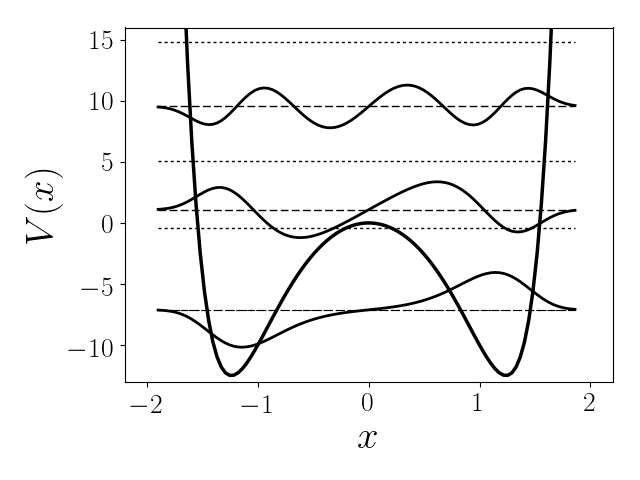}}
\caption{
Left: The three even eigenfunctions (narrow solid lines) found analytically for $k=4$ and $N = 2$, together with the corresponding eigenvalues (dashed lines).
Right: The three odd eigenfunctions (narrow solid lines) found analytically for $k=5$ and $N = 2$, together with the corresponding eigenvalues (dashed lines).
The unsolved eigenvalues are shown in dotted lines. }
\label{figAN2k5}
\end{figure}

For the case with $N=2$, choosing $k=4$, the energy eigenvalues are
$E_{2,0} = -3.74456$, $E_{2,2} = 1.00000$, and $E_{2,4} = 7.74456$.  The corresponding eigenfunctions are plotted in Fig.(\ref{figAN2k5}).

Now, to find the antisymmetric eigenfunctions we set
$f(\beta) = {\rm sinh}(x) \ g(\beta)$, to get the CHE
\begin{eqnarray} \label{ecuacion5.34}
\nonumber \beta (\beta - 1) \frac{d^{2} g}{d \beta^{2}} &+& \left[ -\alpha \beta^{2} + \left( \alpha + 2 \right) \beta - \frac{1}{2}   \right] \frac{d g}{d \beta} \\
&+& \left[\beta \left( \frac{\alpha^{2}}{4} \left( 1 + c \right) - \alpha \right) + \left( \frac{\alpha}{4}+ \frac{E}{2} - \frac{\alpha^{2}}{4}\left( 1 + c\right) + \frac{1}{4} \right) \right]g = 0 \ .
\end{eqnarray}

For $N=0$ we get that $\alpha = 4/(1+k)$ and $E_1 = 6/(1+k) - 1/2$, such that if $k > 11$ we may
find negative energy eigenvalues.  For $N=2$, $\alpha = 12/(1+k)$, if we set $k = 5$
the energy eigenvalues found are
$E_{2,1} = -7.11693$, $E_{2,3} = 1.08119$, and $E_{2,5} = 9.53574$.
The eigenfunctions are plotted in Fig.(\ref{figAN2k5}).

Note that in this case $(E_1-E_0)/E_0=0.0052$, and it is not possible to distinguish these eigenvalue's lines from each other in Fig.(\ref{figAN2k5}) for antisymmetric eigenvalues, implying quasi-degenerate eigenstates.  A similar effect is seen in the symmetric case.


\subsection{The case with $\boldsymbol{k=-1}$}\label{secnew}

As was seen in Section \ref{seckgt0}, the ground state energy diverges as $1/(1+k)$ as $k\to -1$, and this also happens to all higher order even eigenvalues (see eq.(\ref{Ek-n1})). This is a strange behaviour, since it is clear that the potential function has a rather simple functional form for any value of $k$: a single or double well with infinite barriers.
We can see that this is only a characteristic due to the analytical solution procedure, coming from the fact that
the potential strength $V_0$ is also divergent when $k\to -1$.


\subsection{Unclassified QES potentials}\label{secnew2}

Finally, we would like to emphasize that there should be other potential functions which may not be classified form the Lie algebraic methood.\cite{Turbiner}

Indeed, let us consider Schr\"odinger's problem
%
with the potential function
\begin{equation} \label{ecuacion2}
V(x) = \frac{\alpha^2}{2} \cosh^{2}(x) -\frac{3\alpha}{2} \cosh(x) + \frac{\alpha}{\cosh(x)}
\end{equation}

For this problem, the ground state eigenfunction and eigenvalue are given by
\begin{equation} \label{ecuacion9}
\psi = \psi_0 e^{- \alpha \cosh(x) } \cosh(x) \ , \ \ \ \
E = \frac{\alpha^2 - 1}{2}
\end{equation}
while this particular problem does not belong to the class of potentials found using the Lie algebraic method.  Similar potentials may be found which do not belong to that class, leaving space for further developments.


\acknowledgments{ \noindent \noindent This work was partially
supported by CONACYT  179881 grants. PROMEP grants
UGTO-CA-3. This work is part of the collaboration within the
Instituto Avanzado de Cosmolog\'{\i}a. E. Condori-Pozo is supported by a CONACYT graduate fellowship }


\end{document}